\documentclass[twocolumn]{aastex631}
\usepackage{amssymb}
\usepackage{hyperref}
\usepackage{amsmath}

\begin{document}
\title{WASP-69b’s Escaping Envelope is Confined to a Tail Extending at Least Seven Planet Radii}

\author[0000-0003-0298-4667]{Dakotah Tyler}
\author[0000-0003-0967-2893]{Erik A. Petigura}
\affiliation{Department of Physics and Astronomy, University of California,	Los Angeles, CA 90095, USA}

\author[0000-0002-9584-6476]{Antonija Oklop\v{c}i\'{c}}
\affiliation{Anton Pannekoek Institute for Astronomy, University of Amsterdam, Science Park 904, NL-1098 XH Amsterdam, The Netherlands}

\author[0000-0001-6534-6246]{Trevor J. David}
\affiliation{Center for Computational Astrophysics, Flatiron Institute, New York, NY 10010, USA}
%	Abstract
\begin{abstract}
Studying the escaping atmospheres of highly-irradiated exoplanets is critical for understanding the physical mechanisms that shape the demographics of close-in planets. A number of planetary outflows have been observed as excess H/He absorption during/after transit. Such an outflow has been observed for WASP-69b by multiple groups that disagree on the geometry and velocity structure of the outflow. Here, we report the detection of this planet's outflow using Keck/NIRSPEC for the first time. We observed the outflow 1.28 hours after egress until the target set, demonstrating the outflow extends {\em at least} $5.8 \times 10^5$~km or 7.5 planet radii. This detection is significantly longer than previous observations which report an outflow extending $\sim$2.2 planet radii just one year prior. The outflow is blue-shifted by $-$23~km~s$^{-1}$ in the planetary rest frame. We estimate a current mass loss rate of 1~$M_{\oplus}$~Gyr$^{-1}$. Our observations are most consistent with an outflow that is strongly sculpted by ram pressure from the stellar wind. However, potential variability in the outflow could be due to time-varying interactions with the stellar wind or differences in instrumental precision.
\end{abstract}


\section{Introduction}
\label{sec:Intro}
The distribution of exoplanet sizes and orbital distance encodes key aspects of planet formation physics such as rocky core growth, the accretion/loss of gaseous envelopes, orbital migration, and other processes. NASA's Kepler mission revealed that planets between the size of Earth and Neptune occur at a rate of $\sim$1 per star (see, e.g., \citealt{2022AJ....163..179P}). In addition there are notable `deserts' in the Kepler census. One such feature is the `Hot-Neptune Desert', a lack of short-period 2--8~$R_{\oplus}$ planets (\citealt{2011ApJ...727L..44S}). Another is the `Radius Gap,' in which we observe a bimodal size distribution of super-Earths and sub-Neptunes with few planets in between (1.4--2.0~$R_{\oplus}$) (\citealt{2017AJ....154..109F}). 

One theory explains  these `deserts' as a consequence of photoevaporation (see \citealt{2017ApJ...847...29O} and references therein), where  X-ray and Extreme Ultraviolet radiation (10~eV --- 100~keV) heats the the upper layers of planet's H/He envelope which creates a pressure gradient that drives a Parker-like outflow. Given these theories, any real-time observations of envelope loss are illuminating. 

Early works such as \cite{2000ApJ...537..916S} proposed that a transiting exoplanet with sufficiently strong absorption features (such as He~I~10830~\r{A}) in its atmosphere would be superimposed on the stellar flux passing through the planet atmosphere above the limb. The first successful outflow detection was a Lyman-$\alpha$ absorption during the transit of HD 209458b \citep{2003Natur.422..143V}.

While Lyman-$\alpha$ presents a large absorption cross-section, it cannot be observed from the ground because it is blue-ward of the UV cutoff of the Earth's atmosphere. Separately, space-based observations must contend with contamination from the ISM \& geocoronal emission. However, in recent years, observing the He~I~10830~\r{A} absorption line during transit has emerged as another probe of planetary outflows as it is sufficiently populated under favorable conditions (\citealt{2018ApJ...855L..11O}). EUV ($h\nu$ = 26--124~eV) photons ionize He~I and recombination efficiently populates the 2$^{3}$S triplet state which is metastable and can only decay to the ground state through forbidden transitions. While EUV photons populate the 2$^{3}$S state by supplying He ions, FUV ($h\nu$ = 4.8--10~eV) photons act to depopulate it through photo-ionization. Thus, tracing mass loss with this method is sensitive to the host stars SED, specifically the EUV/FUV ratio. Planets orbiting K stars fall into this region for optimal for He~I~10830~\r{A} observability (\citealt{2019ApJ...881..133O}).

The first successful He~I~10830~\r{A} detection was made by \cite{2018Natur.557...68S}, who detected an extended exosphere of the sub-Saturn WASP-107b. Since then, there has been a flurry of mass loss detections using He I 10830~\r{A} for close-in planets orbiting late-type stars.

Another early detection was of WASP-69b, a $1.1~R_{J}$, 0.26~$M_{J}$ planet that orbits a K5 type host once every 3.86 days. \cite{2018Sci...362.1388N} previously observed He~I~10830~\r{A} absorption in the system with the CARMENES spectrograph at Calar Alto Observatory. Over two nights, Nortmann et al reported a relative excess He I absorption of 3.59\% with a net blue-shift of $-3.58$~km~s$^{-1}$. They detected no observable pre-transit absorption, but reported continued absorption for 22~min post-transit which suggests asymmetry in the outflow geometry resembling a comet-like tail.

In another observation of WASP-69b, \cite{2020AJ....159..278V} used an ultra-narrowband filter coupled to a beam-shaping diffuser on the Wide-field Infrared Camera (WIRC) at Palomar Observatory. They reported an excess He~I absorption of 0.498$\pm$0.045$\%$ during transit which is consistent with previous observations when considering their bandpass, however, they did not detect any He~I trailing the planet. We note that the photometric observations do not contain velocity information.

In this paper, we provide a new view of the WASP-69b outflow using Keck/NIRSPEC. Relative to CARMENES, NIRSPEC collects more photons per transit resulting in a higher SNR and allows for a more detailed examination of the tail geometry and velocity structure.

We describe our Keck/NIRSPEC observations in Section \hyperref[sec:Observations]{2}. We explain the data reduction process in Section \hyperref[sec:Data Reduction]{3}. We characterize the strength, time-dependence, and velocity structure of the outflow in section \hyperref[sec:Results]{4}. We estimate a mass-loss rate in section \hyperref[sec:massloss]{5}. We discuss comparisons with previous observations and place our observations of WASP-69b in the broader context of He I detections in section \hyperref[sec:Comparison]{6}. Finally, we conclude in section \hyperref[sec:Conclusion]{7} with our interpretation and potential future work on this system as well as others.

\begin{table}[h]
\centering
\noindent
    \caption{WASP-69b Physical, Orbital, System Parameters}
    \label{tab:table1}
    \begin{tabular}{llcr}
    %\toprule
    \hline
    Parameter & Unit & Value & Source\tablenotemark{*}\\
    \hline
    $T_{eff}$ & [K] & 4700 $\pm$ 50 & CB17\\
    Age &[Gyr] & 7.0 & CB17\\
    \text{Fe/H} & & 0.150 $\pm$ 0.080 & B17\\
    $M_{\star}$ &[$M_{\odot}$] & 0.826 $\pm$ 0.029 & CB17\\
    $R_{\star}$ &[$R_{\odot}$] & 0.813 $\pm$ 0.028 & CB17\\
    $v$ sin($i_{\star}$) &[km s$^{-1}$] & 2.20 $\pm$ 0.40 & CB17\\
    log $g$ &[cm s$^{-2}$] & 4.50 $\pm$ 0.15 & S17\\
    $\gamma$ &[km s$^{-1}$] & $-9.62826$ $\pm$ 0.00023 & A14\\
    \hline
    $M_{p}$ &[$M_{J}$] & 0.2600 $\pm$ 0.0185 & CB17\\
    $R_{p}$ &[$R_{J}$] & 1.057 $\pm$ 0.017 & CB17\\
    $T_{eq}$ &[K] & 963 $\pm$ 18 & CB17\\
    $K_{p}$ &[m s$^{-1}$] & 38.1 $\pm$ 2.4 & CB17\\
    \hline
    $a$ &[au] & 0.04525 $\pm$ 0.00075 & CB17\\
    $a/R_{\star}$& & 12.00 $\pm$ 0.46 & CB17\\
    $T_{c}$ &[BKJD] & 915.8334 $\pm$ 0.0002 & CB17\\
    $P$ &[d] & 3.8681390 $\pm$ 0.0000017 & CB17\\
    $T_{14}$ &[hr] & 2.23 $\pm$ 0.023 & CB17\\
    $i_{p}$ &[deg] & 86.71 $\pm$ 0.20 & CB17\\
    $e$& & 0 & CB17\\
    $b$& & 0.686 $\pm$ 0.023 & CB17\\
    \hline
    %\bottomrule
    \end{tabular}
\tablenotetext{*}{NOTE - Values referenced come from CB17: \cite{2017A&A...608A.135C}, B17: \cite{2017A&A...602A.107B}, S17: \cite{2017AJ....153..136S}, \& A14: \cite{2014MNRAS.445.1114A}.}

\end{table}
%\begin{tablenotes}[flushleft]
%\small
%\end{tablenotes}
\begin{table*}[!ht]

%\noindent
\centering
\caption{Observations}
\label{tab:table2}
%\begin{tabular}{llccccccc}@{\extracolsep{\fill}}
%\begin{tabular*}{\textwidth}{@{}llccccccc@{\extracolsep{\fill}}}
\begin{tabular}{llccccccp{19mm}}
\hline
\text{Target} & \text{Observation} & \text{Date of} & \text{Start Time} &\text{End Time} & \text{Airmass} & \text{$N_{obs}$} & \text{$t_{exp}$} & \text{SNR}$\cdot$\text{pix}$^{-1}$ \\
& & observation & [UT] & [UT] & range &  & $[s]$ & range \\
\hline
HIP102631 & Telluric Std & 2019-07-12 & 09:09 & 09:12 & 1.36--1.34 & 4 & 50 & 50\\
WASP-69 & Pre-Transit & 2019-07-12 & 09:25 & 11:06 & 1.44--1.14 & 34 & 149 & 35--60\\
WASP-69 & In-Transit & 2019-07-12 & 11:09 & 13:32 & 1.13--1.19 & 42 & 149 & 50--62\\
WASP-69 & Post-Transit & 2019-07-12 & 13:36 & 14:47 & 1.20--1.48 & 24 & 149 & 51--58\\
HIP105315 & Telluric Std & 2019-07-12 & 14:59 & 15:06 & 1.6--1.64 & 4 & 100 & 50\\
WASP-69 & Out of Transit & 2022-08-02 & 05:58 & 13:45 & 1.07--3.47 & 74 & 149 & 25--42\\ 
\hline  
\end{tabular}
\end{table*}

\section{Observations}
\label{sec:Observations}
%	Observations
On July 12, 2019 UT, we observed the transit of WASP-69b (see Table \ref{tab:table1} for system parameters) using Keck/NIRSPEC (\citealt{1998SPIE.3354..566M}; \citealt{2018SPIE10702E..0AM}). We used the NIRSPEC-1 filter ($\lambda$ = 9470--11210~\r{A}) and the 0.28 $\times$ 12~arcsec slit, which provides a resolving power of $R$ = 40,000; at the He feature, this corresponds to  $\Delta\lambda$ = 0.3~\r{A}, or $\Delta v$ = 7.5~km~s$^{-1}$. Following previous works, we removed the ``thin''  blocking filter to eliminate CCD fringing in the Y-band. On August 02, 2022 UT we observed WASP-69 again using the same set up and reduction listed throughout this paper. However, due to an time-scheduling error, we did not observe the transit of WASP-69b. However, this did allow us to obtain an out of transit baseline for WASP-69 which we use to further constrain He I variability in the stellar atmosphere. These observations occurred during orbital phase $\phi$~=~$[-0.30,-0.22]$ compared to the transit midpoint at phase of $\phi$~$=0$.

We scheduled our observations using the ephemeris of \cite{2017A&A...608A.135C}. The transit midpoint occurred at 12:20 UT with $\pm$~2.1~min midpoint uncertainty. We used the ABBA nod-dithering technique to remove signal from the background and collected 104 exposures of WASP-69 with integration times of 149~s from UT = 09:25 -- 14:44. The first observation began 108 min before ingress when the target was at airmass~=~1.44; the last observation ended when the target was at airmass~=~1.46. We lost guiding on two frames in the WASP-69 sequence during egress. Additionally, we observed two nearby rapidly rotating A0V stars for telluric calibration at the beginning and end of the WASP-69 observing sequence: HIP102631 (airmass = 1.35, $v$~sin($i$) = 142~km~s$^{-1}$) \& HIP105315 (B9V, airmass = 1.62, $v$~sin($i$) = 200~km~s$^{-1}$). 

\section{Data Reduction}
\label{sec:Data Reduction}
%	Data Reduction
We reduced 104 raw observations using the NIRSPEC Reduction Package, REDSPEC \citep{2015ascl.soft07017K}%
\footnote{https://github.com/Keck-DataReductionPipelines/NIRSPEC-Data-Reduction-Pipeline}.
 We focused on echelle order 70, which encompasses the He I triplet transition lines. In this work, we adopt a vacuum wavelength scale for the He~I triplet transitions (10832.06, 10833.22, \& 10833.31~\r{A}). We established our laboratory wavelength solution using Neon, Argon, Xenon, and Krypton arc lamps provided by NIRSPEC website\footnote{https://www2.keck.hawaii.edu/inst/nirspec/lines.html}. 

We derived an initial wavelength solution and reduced each AB and BA nod pair into single, combined 1D frames. We used the suggested SNR measurement formula for the 2018 NIRSPEC update and report a SNR of $\approx$~55 per reduced pixel.  Due to the target hopping off of the slit mentioned in the previous section, there is a small gap of missing data in our time series which occurs just as the the planet begins egress.

We used REDSPEC to identify bad pixels and cosmic ray hits and interpolated over them using a cubic spline.

\begin{figure}[!h]
	\includegraphics[width=1\columnwidth]{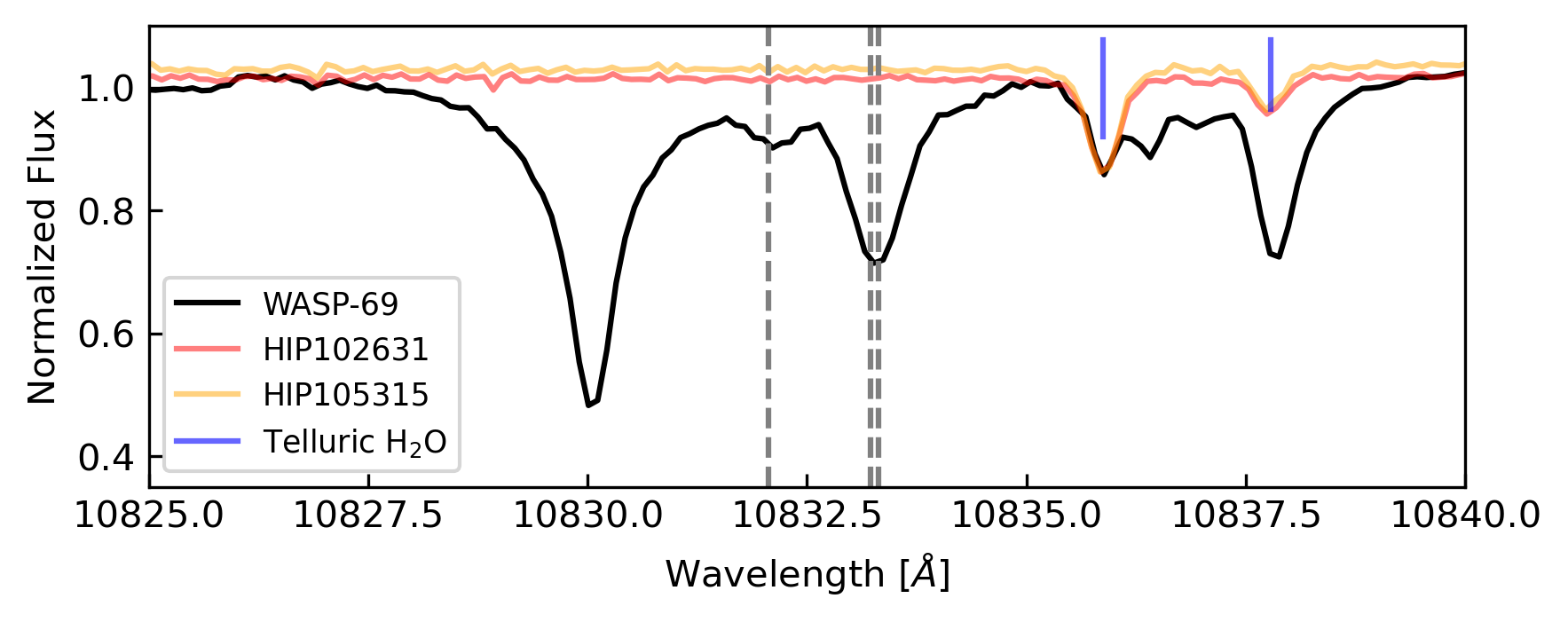}
	\caption{The two calibration stars, HIP102631 (red) \& HIP105315 (orange), are shown here with a slight offset. The nearby telluric absorption features at 10835.9 \& 10837.8~\r{A} are marked with blue. A representative WASP-69 exposure is shown in black and conveys the spectral region, instrument resolution, and signal-to-noise ratio. The three dashed gray lines indicate the He~I~10830~\r{A} triplet vacuum rest wavelengths. The main core of the profile comes from the 10833.22 \& 10833.31~\r{A} features which are blended. The weaker singlet is located at 10832.06~\r{A}.}
	\label{fig:tellurics}
\end{figure}

We used the calibration star spectra to identify telluric features in the target spectrum and inspect for contamination of the He I profile. We identify nearby H$_{2}$O absorption lines at 10835.9 and 10837.8~\r{A}, but they do not interfere with the absorbing He I band as shown in Figure~\ref{fig:tellurics}. 

We shifted the spectra to the stellar rest frame by cross-correlating each spectrum with a PHOENIX stellar template model (\citealt{2013A&A...553A...6H}) with the following parameters: $T_\mathrm{eff}$ = 4700 K; log($g$) = 4.5; [M/H] = 0.0). We masked out the He~I absorbing region (10830--10835~\r{A}) as well as any telluric features so they do not influence our spectral registration. 

 \begin{figure*}[!ht]
 \centering
\includegraphics[width=\textwidth]{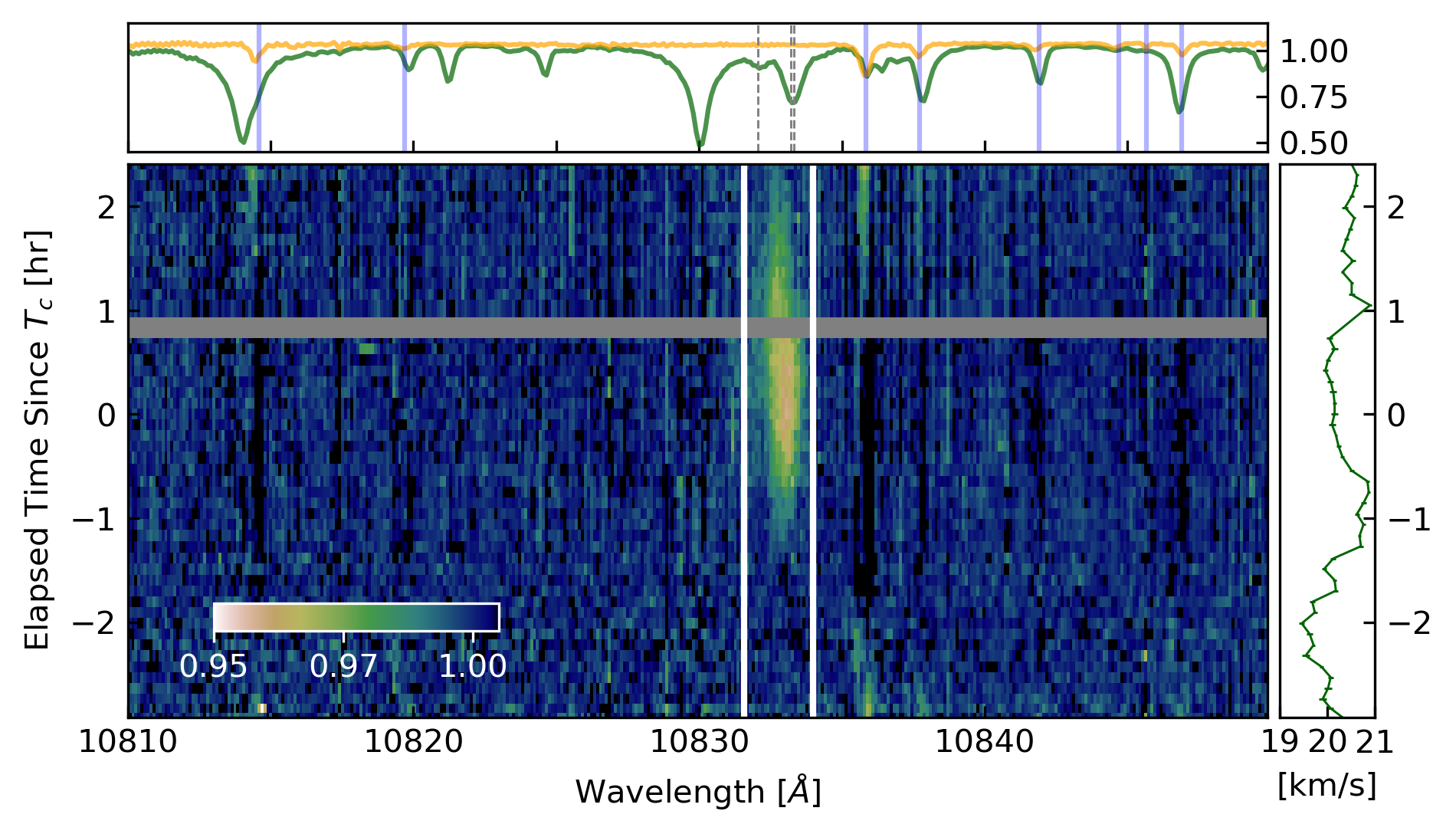}
	\caption{Colors in the main panel the depth of each normalized spectrum {\em relative} to a master template made up of all pre-transit observations, i.e. 1.0 corresponds no excess absorption. Missing data is gray and the vertical white lines encompass the defined He~I bandwidth (see Section \ref{sec:Observations} for further details). The top panel shows the green master spectrum for WASP-69 and the orange HIP105315 telluric standard spectrum for which we have applied an arbitrary vertical offset for clarity. Telluric absorption features are indicated with blue vertical lines and coincide with the transient vertical artifacts in the plot. The right panel shows the velocity shift between the PHOENIX model and each frame.}
	\label{fig:modified_plots}
\end{figure*}

\begin{figure}[!h]
\includegraphics[width=1\columnwidth]{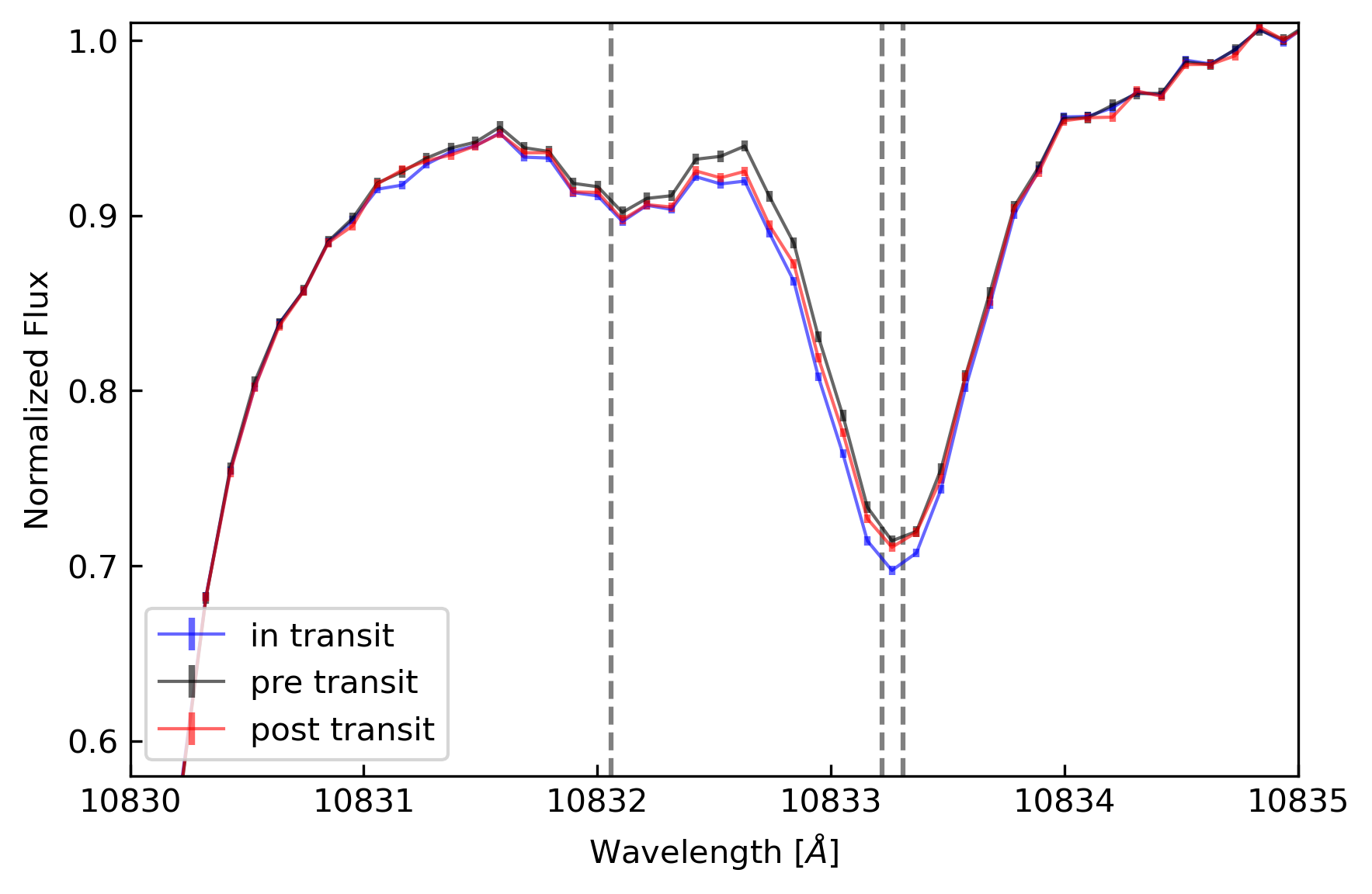}
\caption{The He~I feature is deeper in-transit (blue spectrum) and post-transit (red spectrum) compared to pre-transit (black spectrum). Gray dashed lines represent the central rest wavelengths in the stellar rest frame.}
\label{fig:summed_IT_OOT}
\end{figure}

We fit the resulting cross-correlation function peaks with Gaussian profiles and determined velocity shifts for each spectrum. These shifts ranged from 19--21~km~s$^{-1}$ and included barycentric Doppler shifts and instrument variability. With our observations now shifted into the stellar rest frame, we co-added the spectra into three distinct bins: pre-transit, in-transit, and post-transit (see Table \ref{tab:table2}). We computed the average spectrum for each bin and plot the composite spectra and corresponding shifts in Figure~\ref{fig:modified_plots}. To interpret the outflow with respect to the WASP-69b, we now calculate the projected velocity of the planet and shift the spectra into the planetary rest frame. 

 \begin{figure*}[!ht]
 \centering
\includegraphics[width=\textwidth]{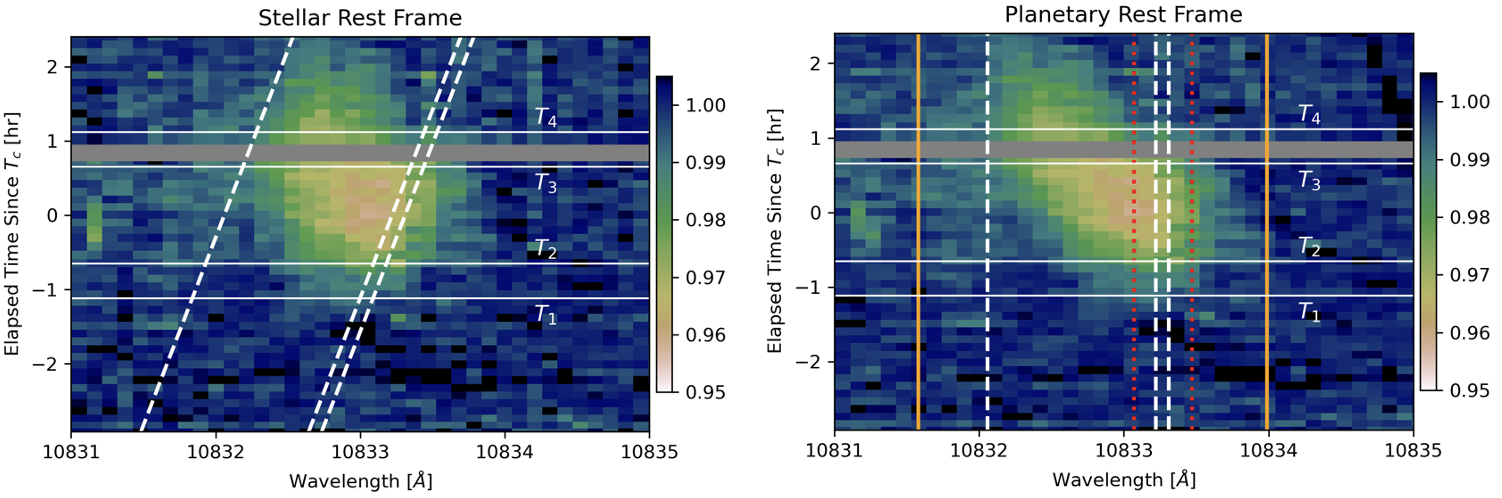}
	\caption{Left: Same as Figure~\ref{fig:modified_plots} but detailing the He~I triplet. $T_{1}$, $T_{2}$, $T_{3}$, and $T_{4}$ label the four transit contact times. The three white dashed lines represent the expected motion of the He~I transition lines in the stellar rest frame due to the predicted planetary velocity. Right: Same as left, but shifted into the planetary rest frame. Two separate Helium bandpasses used to sum up the Helium light curve are plotted as vertical orange and red-dotted lines. The orange vertical lines represent our preferred bandpass. The red-dotted lines are comparable to the bandpass used in other data sets.}
	\label{fig:waterfall}
\end{figure*}

\section{Results}
\label{sec:Results}
%	Data Analysis

In Figure~\ref{fig:summed_IT_OOT}, we show the normalized pre-transit, in-transit, and post-transit absorption profiles in the He I band with respect to the central rest wavelength positions of the He I triplet. Excess absorption is clearly visible in-transit and post-transit. The difference in absorption depth and width of the bluer wing of the profile is evident.

In Figure~\ref{fig:waterfall}, we plot the relative absorption intensity in both the stellar and planetary rest frames. The first through fourth contact points are shown as well as the central positions of the He~I~10830~\r{A} triplet. We report an average in transit absorption depth of 2.7\%~$\pm$~0.4\%. We estimated the uncertainty over the He~I~10830~\r{A} bandwidth [10831.58--10833.99~\r{A}] by measuring the standard deviation of the average normalized intensity in neighboring bins of equal extent in time and wavelength. This dispersion captures the photon-counting statistics and instrumental errors associated with variations in wavelength solution and line profile. Assuming a spherical outflow geometry, the relative absorption depth tells us about the extent to which the extended exosphere is enveloping the planet.

We computed the velocity of the He~I triplet by modeling the profile of the line with 3 Gaussians, each corresponding to the three central wavelengths of the metastable He~I transition. We fixed the relative heights and wavelength between the transitions and allowed two parameters to vary: the total depth and the central wavelength. During the in-transit observations, we detect a net blue-shift of $-5.9$ $\pm$ 1.0~km~s$^{-1}$. This radial velocity shift in the absorption profile can trace the bulk velocity structure of the outflowing He~I.

\begin{figure*}[!ht]
 \centering
\includegraphics[width=\textwidth]{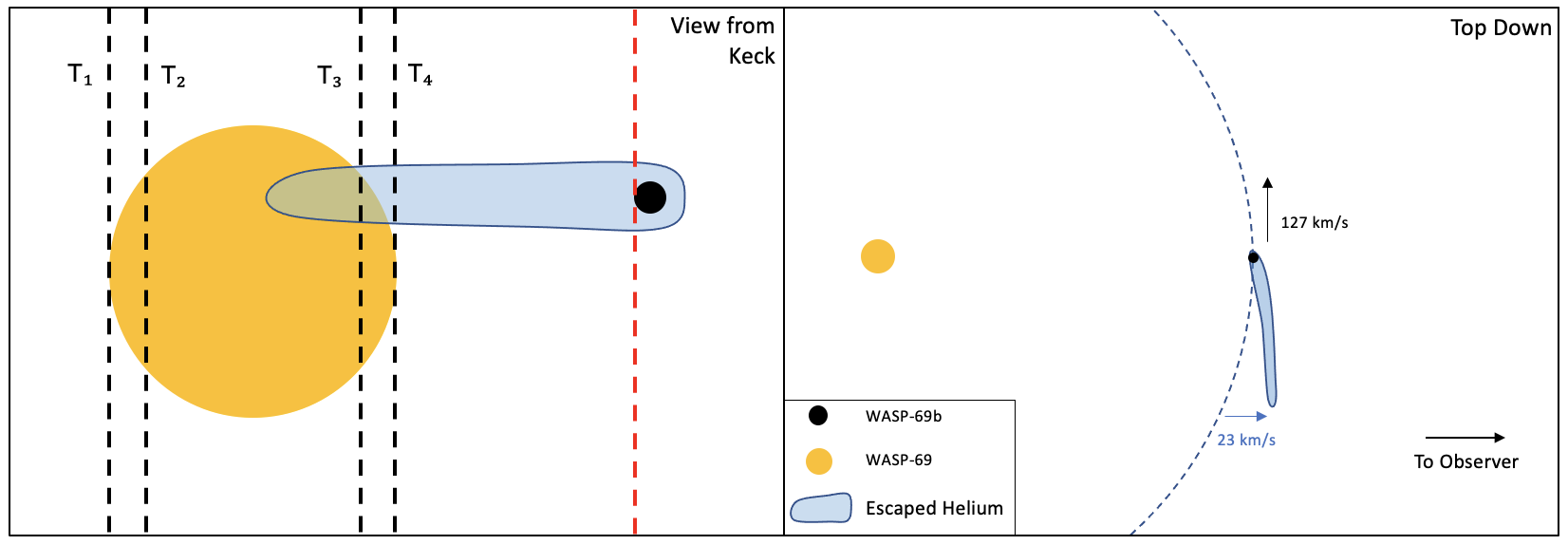}
	\caption{Transit chord and top down view of the WASP-69 system presented to scale. Left: Transit chord view from Keck. The four contact points, $T_{1}$, $T_{2}$, $T_{3}$, and $T_{4}$ are represented with vertical black dashed lines and the absorbing He~I is light blue. The red dashed line represents the final predicted position of the planet corresponding to the last observation in the spectral time series after traveling over 7 $R_{p}$ (1.28~hrs) beyond the disk of the star from the perspective of the observer. Right: Top down view of the system. The He~I tail can be seen accelerating towards the observer on the lower right of the panel.} 
	\label{fig:sketch}
\end{figure*}

We can see that the excess absorption continues for the entire post-transit sequence (1.28 hours). Furthermore, the absorbing He~I is blue-shifted for the duration of the transit and accelerates away from the star. This excess He~I absorption post-transit is consistent with a ``comet-like" tail of helium trailing the planet and continuing to absorb stellar photons while being accelerated away from the planet and towards the observer due to stellar winds. A simple diagram of this transit chord can be seen in Figure~\ref{fig:sketch}. As the planet has traveled over 7~$R_{p}$ beyond the disk, a continually thinning column of He~I is still detected.

These tails are shaped by stellar winds which interact with the planetary outflow and redirect material around the planet, radially away from the star (\citealt{2022ApJ...926..226M}). \cite{2019ApJ...873...89M} showed that the degree to which the planetary outflow is shaped is a function of orbital velocity, intrinsic planetary wind velocity, and stellar wind strength - which can be variable itself. While weak stellar winds cannot confine the planetary outflow, a strong stellar wind can suppress the outflow and redirect the majority of the out-gassed material into the tail. Observational evidence of this was first seen in Lyman-$\alpha$ detections, the most extreme for the hot-Neptune GJ436b (\citealt{2017A&A...605L...7L}) which has a hydrogen tail that continues absorbing for several hours after the optical transit ends. 

\begin{figure}[!b]
	\includegraphics[width=1\columnwidth]{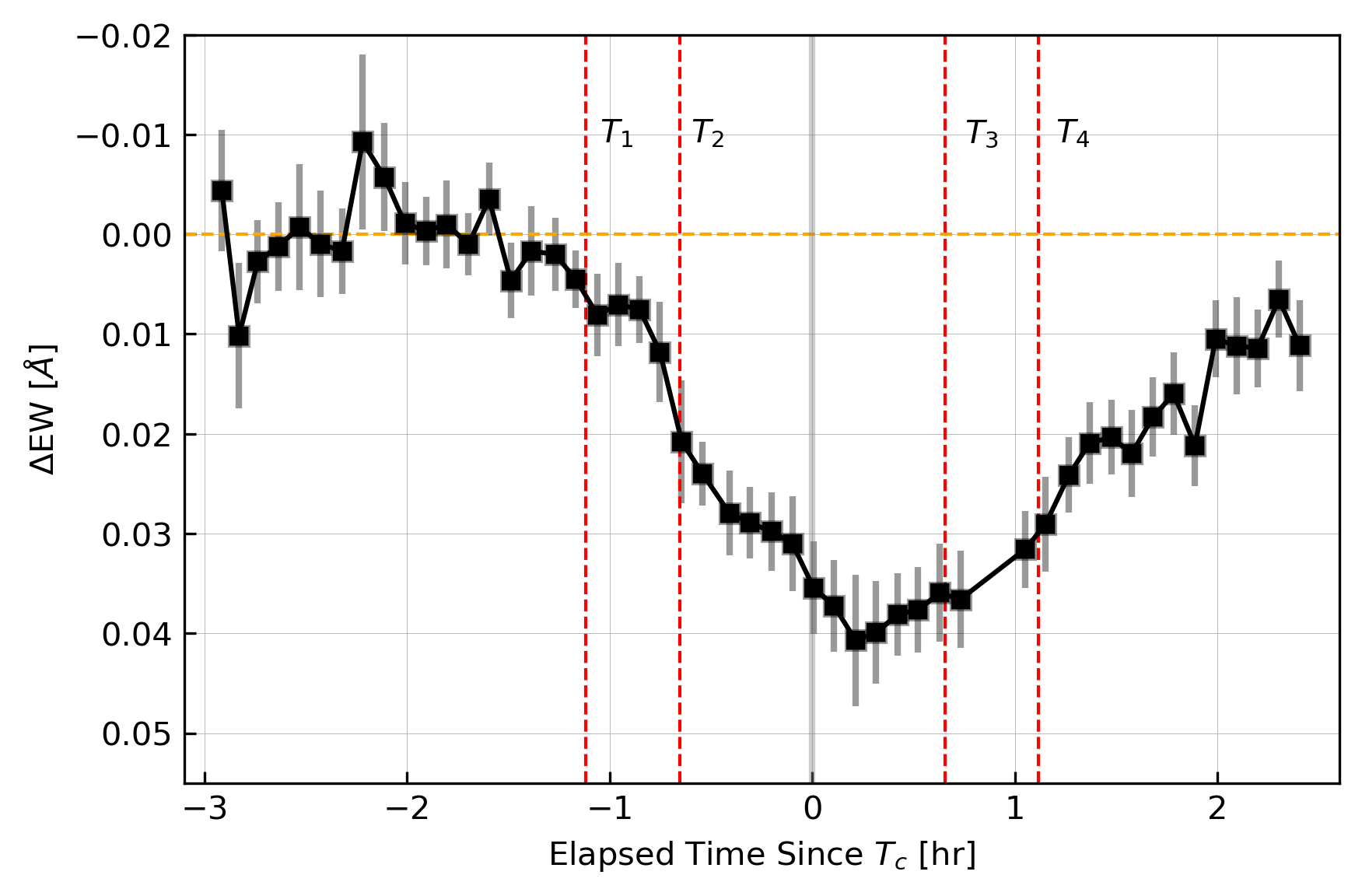}
	\caption{Equivalent Width variations over the course of the night. The red dashed lines represent the four contact points and the horizontal orange line shows the pre-transit baseline which is not recovered. The maximal depth is slightly delayed compared to mid-transit, as predicted by modeling strong stellar wind interactions that shape planetary outflows.}
	\label{fig:EW}
\end{figure}

The time asymmetry of the absorption profile is consistent with interactions between planetary outflows and strong stellar winds (\citealt{2022ApJ...926..226M}). In general, a planetary outflow will form a bow shock at the point where the outflow ram pressure $\rho$$v^{2}$ equals the ram pressure of the stellar wind.  Multiple groups have shown that for sufficiently strong stellar winds, the planetary outflow can be suppressed below the sonic surface and completely redirected around the planet and toward the tail (\citealt{2022ApJ...926..226M}; \citealt{2021ApJ...914...99W}).

To probe the extent of the absorption, we computed the equivalent width for the spectral time series. We defined the He~I absorbing band [10831.58--10833.99~\r{A}] to include the full wavelength range where excess absorption is detected (Figure~\ref{fig:modified_plots}).   

We report a maximum EW just after mid-transit of 40.7~m\r{A} $\pm$ 6.8~m\r{A}. During the transit ($T_{14}$), we find an average EW of 27.8 m\r{A} $\pm$ 2.5~m\r{A}. As shown in Figure~\ref{fig:EW}, the post-transit equivalent width never returns to the pre-transit baseline. We compute an average EW of 15.4~m\r{A} $\pm$ 3.9~m\r{A} during the post-transit phase. Here the time asymmetry can be seen clearly and in the last observation (1.28~hrs after $T_{4}$) we report an EW~5\% higher than the pre-transit average.

Simulations show that stellar winds can not only cause time asymmetry in the relative equivalent width absorption of the He~I feature, but also shift the peak of maximal absorption relative to the optical point of conjunction (\citealt{2022ApJ...926..226M}; \citealt{2021ApJ...914...99W}). After the peak absorption, the decrease in absorption transitions in to a more gradual return to the pre-transit baseline that can take several hours. 

Assuming a circular orbit, we calculate WASP-69b's orbital velocity $v_\mathrm{orb}$ = 2$\pi~a_{p}/P$ = 127.3~$\pm$~1.5~km~s$^{-1}$. In the 1.28~hours of post-transit observations that we detect continued absorption, the planet travels $\sim$~5.9~$\times$~10$^{5}$~km. Assuming all of the helium we are seeing is confined to a partial annulus at the orbital separation of the planet, we set a minimum limit on the length of the He~I tail $\geq$~7.5~$R_{p}$. The Roche lobe for WASP-69b is located at $\approx$~2.7~$R_{p}$ but the tail extends well beyond that limit and is unbound from the gravitational influence of the planet. 

\section{Mass-Loss Rate Estimate}
\label{sec:massloss}

One of the main goals for tracing He~I is to quantitatively measure mass loss rates in real-time. A detailed model of this outflow would include effects like Coriolis force and advection coupled with a 3D radiative transfer scheme and is beyond the scope of this work (see, e.g., \citealt{2021ApJ...914...98W}). Here, we offer two estimates of the mass-loss rate of WASP-69b: 1) An order-of-magnitude estimate following the method of \cite{2022AJ....163...67Z} and 2) an estimate assuming a 1D Parker-like outflow.

\subsection{Order of Magnitude}
We assumed that most of the planetary outflow is optically thin, which is consistent with the  weakness of the singlet centered near 10832~\r{A}. We then take the optical depth from star to observer to be
\begin{equation}
\tau (\lambda) = n_{He} {\sigma_{\lambda}} P (\lambda),
\end{equation}
where $n_{He}$ is the column density of metastable helium atoms, $P(\lambda)$ is the line profile with $\int_{-\infty}^{+\infty} P(\lambda)d\lambda~=~1$, and the absorption cross-section is ${\sigma_{\lambda}}$ $\equiv$ $(\pi e^2g_{l}f_{l}\lambda_{0}^2)/(m_{e}c^2)$. Here, $e$ is the electron charge, $m_{e}$ is the electron mass, $g_{l}$ is the statistical weight for the lower level, $f_{l}$ is the oscillator strengths of the three lines (0.059, 0.179, and 0.299, respectively; \citealt{NIST_ASD}), and $\lambda_{0}$ is the rest wavelength of the absorbing spectral line.  Assuming the optically thin limit where $1 - e^{-\tau} \approx \tau$, we integrate over $\lambda$ and obtain the standard equation for equivalent width:
\begin{equation}
W_{\lambda} = N \sigma_{\lambda}
\end{equation}
Defining the average equivalent width $W_{avg}$, we can solve for the total number of metastable helium atoms $N_{He_{3}^{2}S}$ by integrating over and dividing by the cross-sectional area of the star:
\begin{equation}
\begin{split}
%\begin{flalign*}
W_{avg} & = \frac{1}{\iint dS} \iint W_{\lambda} dS \\
\iint W_{\lambda} dS & = W_{avg} \iint dS \\
N_{He_{3}^{2}S} \sigma_{\lambda} & = W_{avg} \pi R_{\ast}^{2} \\
N_{He_{3}^{2}S} & =\frac{R_{\ast}^{2}m_{e}c^{2}}{e^{2}g_{l}f_{l}\lambda_{0}^{2}}W_{avg}\\
\end{split}
%\end{flalign*}
\end{equation}

Using our calculated EW measurement for $W_{avg}$ we calculate $N_{He_{3}^{2}S}=3.1\times10^{32}$ which we can convert into a total amount of helium assuming metastable helium comprises $10^{-6}$ of total helium nuclei. This fraction assumes an optimistic case that applies to early K-type stars (as shown in \citealt{2019ApJ...881..133O}). Assuming primordial mass and number composition ratios for helium to hydrogen, 3:1 and 9:1 respectively, we estimate the total mass of the planetary outflow in helium and hydrogen to be $m_{tot}=1\times 10^{16}~g$. With this mass estimate, we can estimate a total mass loss rate $m_\mathrm{tot}/ \tau$, where the replenishment lifetime for observable He~I atoms crossing the stellar disk is $\tau$ = $R_{\star}/c_{s}$, where $c_{s}$ is the sound speed. We adopt 10~km~s$^{-1}$ which is consistent with a typical sound speed for planets with these outflows. Our order-of-magnitude mass-loss estimate is $\dot{M} =  m_\mathrm{tot}/ \tau= 1~M_{\oplus}$~Gyr$^{-1}$ or 1.8~$\times$~10$^{11}$g~s$^{-1}$.

\subsection{1-D Parker Model}
As an alternative to our order of magnitude estimate, we use \texttt{p-winds} \citep{2022A&A...659A..62D} to estimate a mass-loss rate by fitting a Parker wind model to our observations. \texttt{p-winds} is an open-source code that implements the 1D model described by \cite{2018ApJ...855L..11O} and \cite{2020A&A...636A..13L} and takes as input the stellar XUV spectrum and the observed Helium transmission spectrum.

As noted, the 1D model is insufficient for modeling outflows with significant asymmetry, however, it is a useful point of comparison to our order of magnitude model and to previous observations. Since the XUV spectrum of WASP-69 is not known, we obtained the MUSCLES XUV spectrum ($\lambda = 10 - 1000$~\r{A}, $h\nu = 1200 - 12$~eV)  of similar star HD85512 (K5) as a proxy for WASP-69. The MUSCLES spectra are observed XUV fluxes at Earth which are then scaled to the appropriate semi-major axis for WASP-69b.

With \texttt{p-winds}, we derived a sound speed of 9~km~s$^{-1}$ at the sonic point which occurs at $\sim3$~$R_{p}$. For a Parker-type hydrodynamic wind, this effectively represents the regime where the pressure-driven flow is no longer being controlled by the planet’s gravity. We derived a temperature in this part of the thermosphere of $9900~\pm~900$~K. We found the total fraction of helium in the metastable state to be $5.4\times10^{-6}$, in agreement with the optimistic case for the environment around a K-type star (\citealt{2019ApJ...881..133O}). The estimated mass-loss rate for the 1D model is $m_\mathrm{tot}/ \tau$ = 1~$M_{\oplus}$~Gyr$^{-1}$ (2.0~$\times$~10$^{11}$~g~s$^{-1}$).

We note as a sanity check that the assumptions made in our order of magnitude estimate are comparable with the \texttt{pwinds} 1D results. Given the difference in these approaches, the agreement is better than expected and we trust both methods at the order of magnitude level.

Assuming a constant orbital distance and stellar output over time, the current mass loss rate suggests that WASP-69b has lost $\sim$~7~$M_{\oplus}$ over the course of the system's $\sim$~7~Gyr lifetime. At the current rate, WASP-69b (92~$M_{\oplus}$) is not at risk of losing its envelope before the end of the lifetime of the system. 

\begin{figure}[!h]
	\includegraphics[width=1\columnwidth]{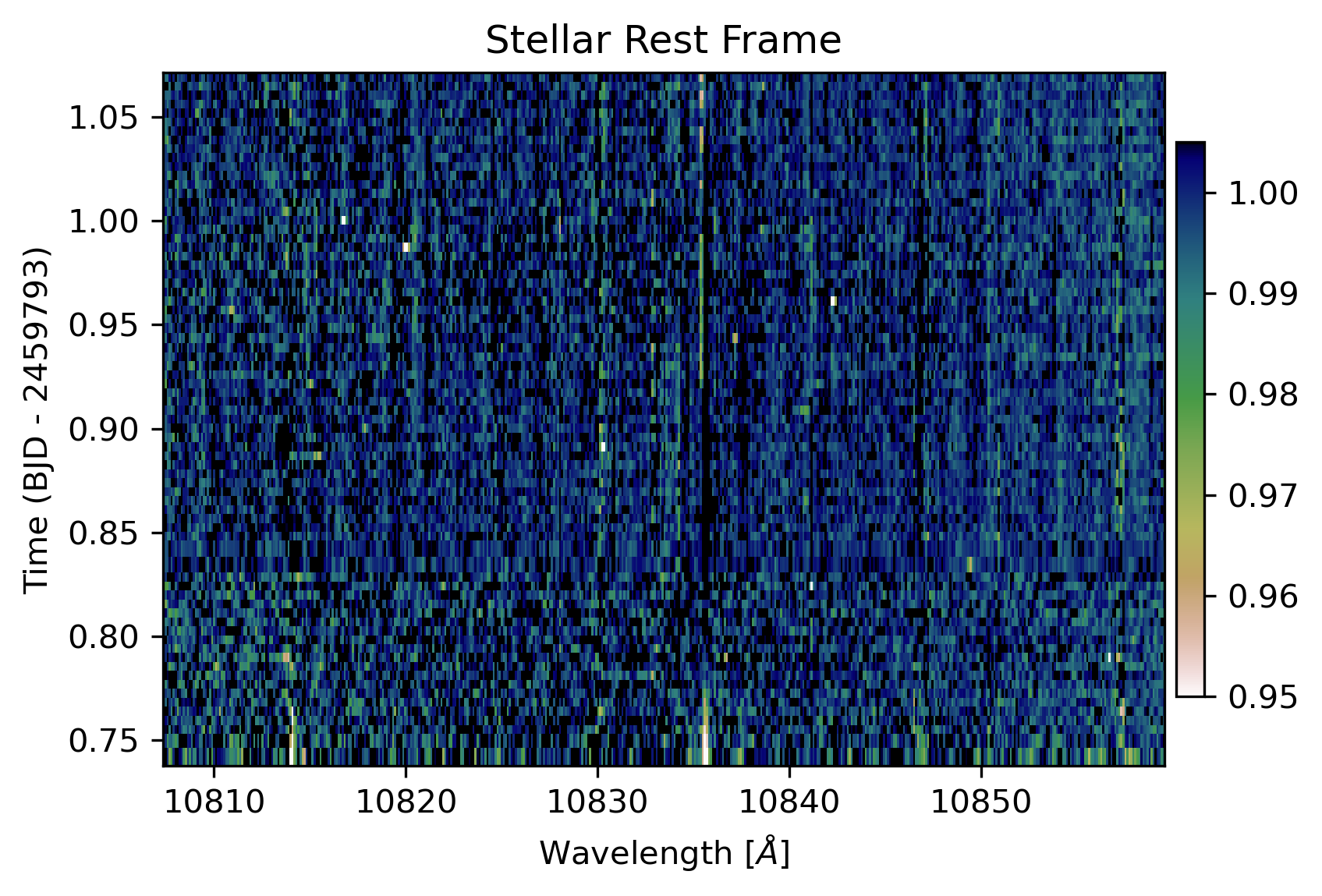}
	\caption{WASP-69 out of transit spectral time series for 2022-08-02. We measure a variability over the He~I triplet band of 0.29$\%$ and report an average equivalent width of $0.18~\pm~0.03~$\r{A}}
	\label{fig:stellar_notransit.png}
\end{figure}

In the 2022 out-of-transit spectrum for WASP-69, which can be seen in Figure~\ref{fig:stellar_notransit.png}, we measured a variability over the main core of the stellar He~I triplet of 0.29$\%$ throughout the night. This is consistent with work from \cite{2000astro.ph.12186C} which showed that inter-night variations of relative He~I absorption strength are $\leq$ 0.25$\%$. These variations fall below our noise floor. We report a pre-transit EW average in 2019 of 0.23~$\pm$~0.03\r{A}. We compare this to the averaged EW we measured for the 2022 WASP-69 out of transit time series of 0.18~$\pm$~0.03\r{A}. The absolute difference in EW between both out of transit epochs is 0.05~$\pm$~0.06~\r{A}. While we don't see any evidence for significant variation within the night on 2022, this does not rule out more long-term variability in the stellar He~I line, or that the pre-transit EW from 2019 was partially contaminated by the "leading arm" of the planetary outflow.

\section{Comparison to Previous Observations}
\label{sec:Comparison}

\begin{figure}[t]
	\includegraphics[width=1\columnwidth]{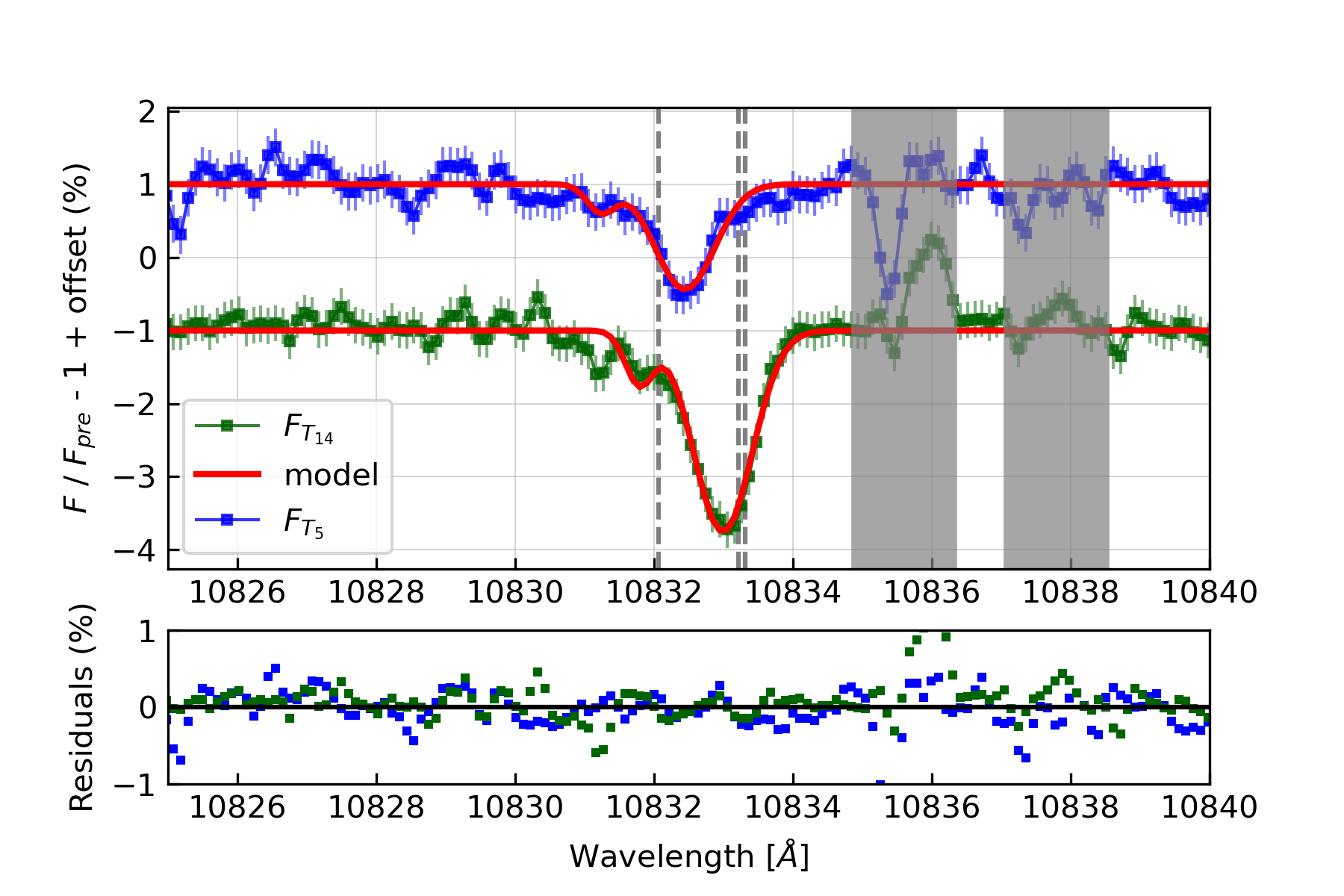}
	\caption{The green points show the averaged in-transit transmission spectrum of He~I 10830~\r{A} absorption. The blue points show the averaged post-transit transmission spectrum which is significantly blue shifted ($-23.3$~$\pm$~1.0~km~s$^{-1})$. The grey vertical masks cover nearby telluric features and the residuals of the models are shown below.}
	\label{fig:transmission}
\end{figure}

\begin{figure}[!h]
    \includegraphics[width=1\columnwidth]{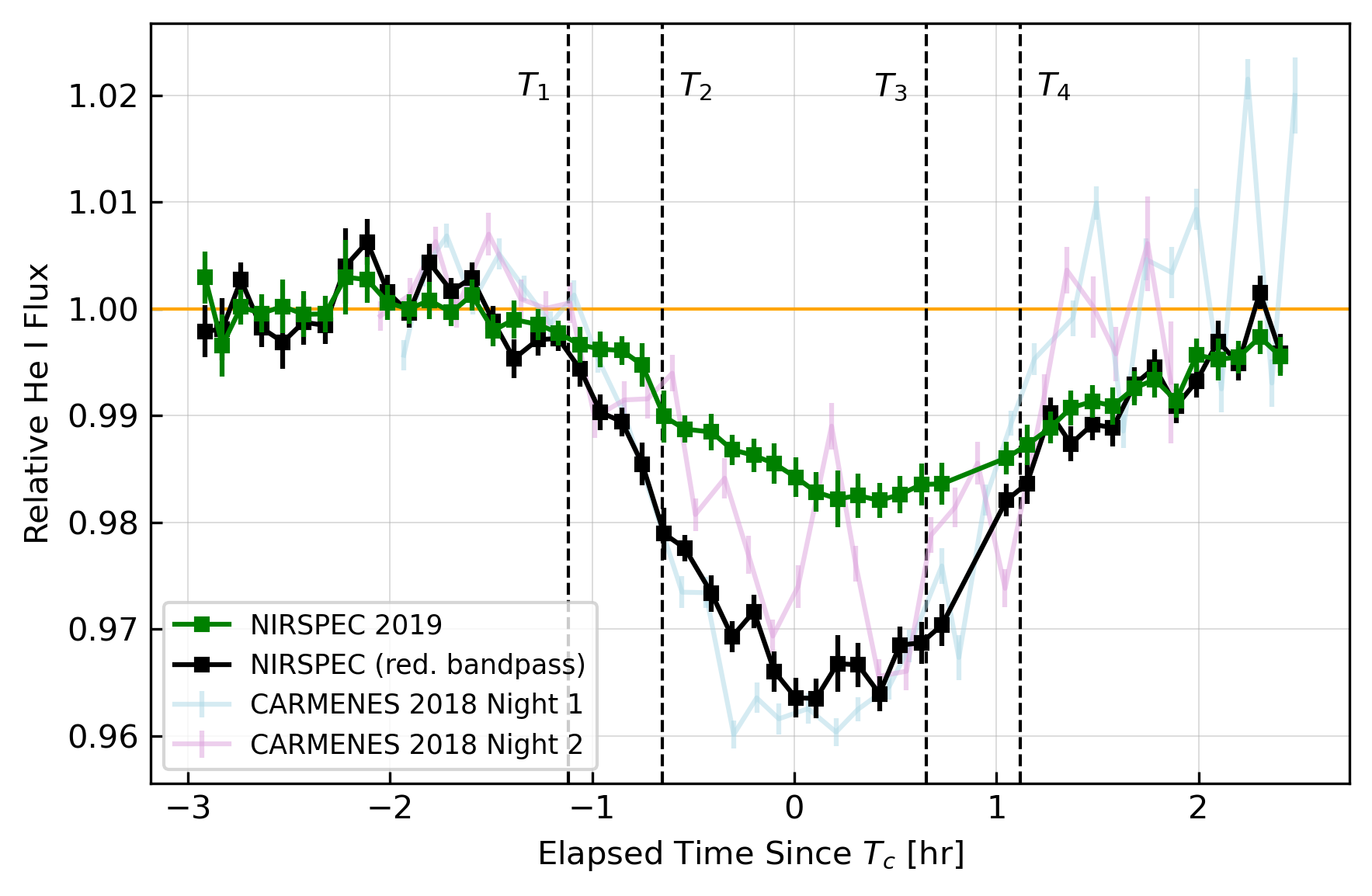}
    \caption{Helium light curves plotted from this work and two nights of CARMENES observations from \cite{2018Sci...362.1388N}. The NIRSPEC observations never return to the baseline post-egress (green). The CARMENES observations do return to the baseline helium absorption level, at least within the achieved precision. Note that for the CARMENES data the point-to-point scatter is significantly higher than the formal uncertainties, which suggests systematic errors or are simply the effects of a reduced signal with lower SNR. Plotted in black is the NIRSPEC data, but with a reduced bandpass that is consistent with the He~I bandpass of \cite{2018Sci...362.1388N} [10833.07--10833.47~\r{A}] for a better comparison. Our preferred bandpass (green points) [10831.58--10833.99~\r{A}] allows us to detect lower levels of He~I absorption in the post-transit sequence due to the high SNR of NIRSPEC. Both Helium bandpasses can be seen plotted in Figure~\ref{fig:waterfall}}
    \label{fig:comparison}
\end{figure}

Our in-transit analysis is consistent with previous CARMENES observations made by \cite{2018Sci...362.1388N} who reported a blue shift of $-3.58$~$\pm$~0.23~km~s$^{-1}$ and 3.59~$\pm$~0.19\% excess absorption (compared to our values $-5.9$~$\pm$~1.0~km~s$^{-1}$ and 2.7~$\pm$~0.4\% excess absorption). We note here that our values are measured from an average of the entire transit ($T_{14}$) whereas the CARMENES values come from the average during $T_{23}$ when the signal is the strongest. Although our measured values are in agreement, this difference does result in a lower reported absorption during our analysis.  However, our post-transit results are inconsistent with their measurements. \cite{2018Sci...362.1388N} reported an average post-transit absorption of 0.5\% and net blue-shift of $-10.7$~$\pm$~1.0~km~s$^{-1}$. During the same phase, we detected an average post-transit absorption of 1.5 $\pm$ 0.2\% and a net blue-shift of $-23.3$~$\pm$~0.9~km~s$^{-1}$ (Figure~\ref{fig:transmission}). This is a significant difference and can be seen in the Helium light curve comparison for the post-transit sequence in Figure~\ref{fig:comparison}. The scatter in the CARMENES data is much higher than the formal uncertainties. Note that this visualization is highly dependent on choice of bandpass for Helium absorption.

\begin{figure}[!h]
	\includegraphics[width=1\columnwidth]{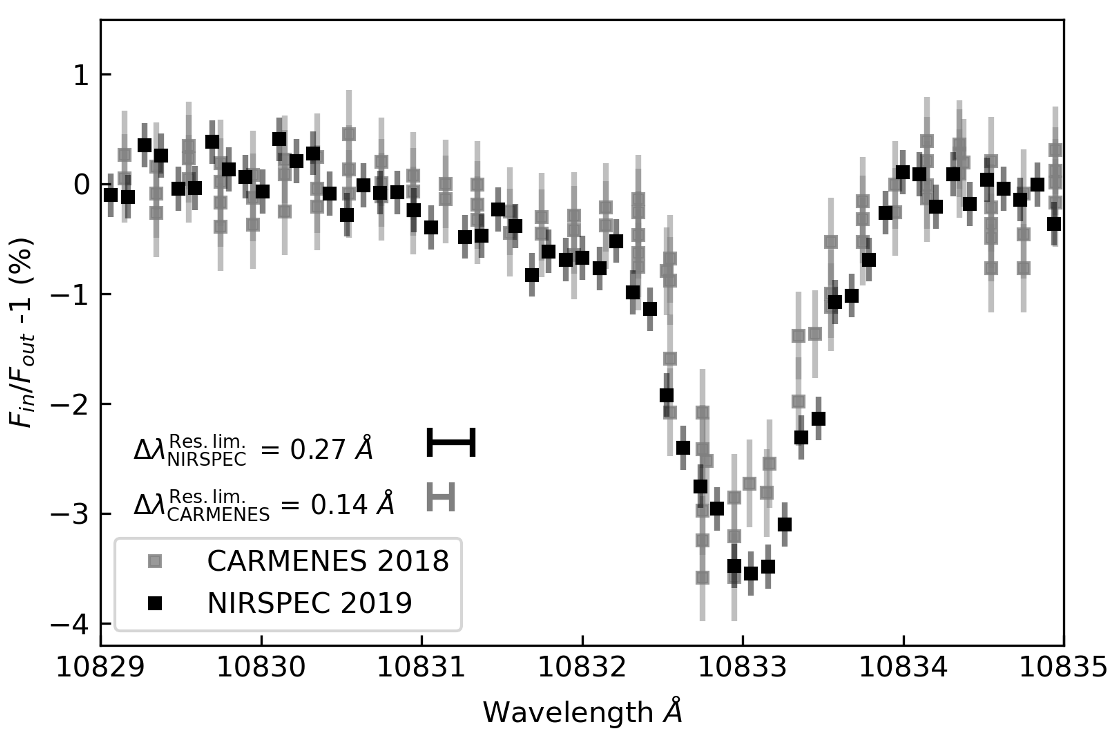}
	\caption{He~I~10830 \r{A} absorption during $T_{2}-T_{3}$ from \cite{2018Sci...362.1388N} and this work. For reference, the resolving limits are plotted for both instruments. The He~I~10830~\r{A} line is clearly resolved in both cases. Thus, we can rule out instrument resolution dependent variability between these data sets.}
	\label{fig:Maintransit}
\end{figure}

CARMENES ($R$ = 80,000) has a higher resolution than NIRSPEC ($R$ = 40,000), but it is on a 3.5m-telescope and collected 8 times fewer photons per transit. The 10m Keck II dish allowed us to get significantly higher SNR per~\r{A} ($\sim$380 compared to $\sim$70 for the CARMENES observations) which likely explains the reported differences between our measurements. In Figure~\ref{fig:Maintransit} we plot our 2019 He~I transmission spectrum with the two nights of data provided by \cite{2018Sci...362.1388N} during $T_{23}$. The He~I~10830~\r{A} triplet is well-resolved in both sets of observations. When the signal is the strongest, there is less discrepancy between the data sets. However, when the He~I becomes more diffuse, the SNR differences between the two instruments is the likely cause for the observed variability in the post-transit tail length. \cite{2020AJ....159..278V} reported comparable SNR per pixel as the CARMENES data set so the lack of significant post-transit He~I absorption in the WIRC observations can also be explained by lower SNR levels.

The previous CARMENES observations from \cite{2018Sci...362.1388N} were used in a 3D Hydrodynamics model by \cite{2021ApJ...914...98W}. They reported a mass loss rate $\sim$ 0.5~$M_{\oplus}$~Gyr$^{-1}$ for WASP-69b but no helium tail. Those results seem consistent with the data set they used. Similarly, \cite{2020AJ....159..278V} report a mass loss rate of $\sim$~0.2~$M_{\oplus}$~Gyr$^{-1}$ which they computed using the same \texttt{p-winds} 1D model that assumes a symmetrical planetary outflow (\citealt{2018ApJ...855L..11O}). Our mass-loss rate estimate of 1.0~$M_{\oplus}$~Gyr$^{-1}$ is higher than both previous results but within an order of magnitude. We require hydrodynamic 3D-modeling for the most accurate mass-loss estimate for WASP-69b.

\cite{2021AJ....162..284S} observed the WASP-107 system with NIRSPEC and reported similar results to these. WASP-107b ($M_{p}$ = 0.1~$M_{J}$, $R_{p}$ = 0.9~$R_{J}$) has a 5.7~d orbital period around a 0.7~$M_{\odot}$ host star. They reported continued He~I absorption for over an hour post-transit which corresponds to a He~I tail length of 5.0 $\times$ $10^{5}$~km. At the time, this was the most dramatic post-transit He~I tail observed. At the lower limit of 5.8 $\times$ $10^{5}$~km, the tail of WASP-69b is at least that long.

Another common indicator for the extent of He~I abundance is the equivalent height of the absorbing atmosphere (\citealt{2018Sci...362.1388N}). Equivalent height is a useful parameter to characterize the radius of the planet in the He~I 10830~\r{A} profile. Equivalent height is defined as $\delta_{R_{p}} = (\Delta d R_{\star}^{2} + R_{p}^{2})^{1/2} - R_{p}$, where $\Delta d$ is the transit depth in He~I. We can compare this equivalent height to the lower atmospheric scale height $H_{eq} = (k_{B}T_{eq})/(\mu g)$, where $\mu$ = 2.3 $\times$ $m_{H}$ which is consistent with a solar H-He composition, to determine how extended the absorbing atmosphere is beyond the expected opaque limit of the atmosphere of a planet with a given temperature. 

 \begin{figure}[h]
	\includegraphics[width=1\columnwidth]{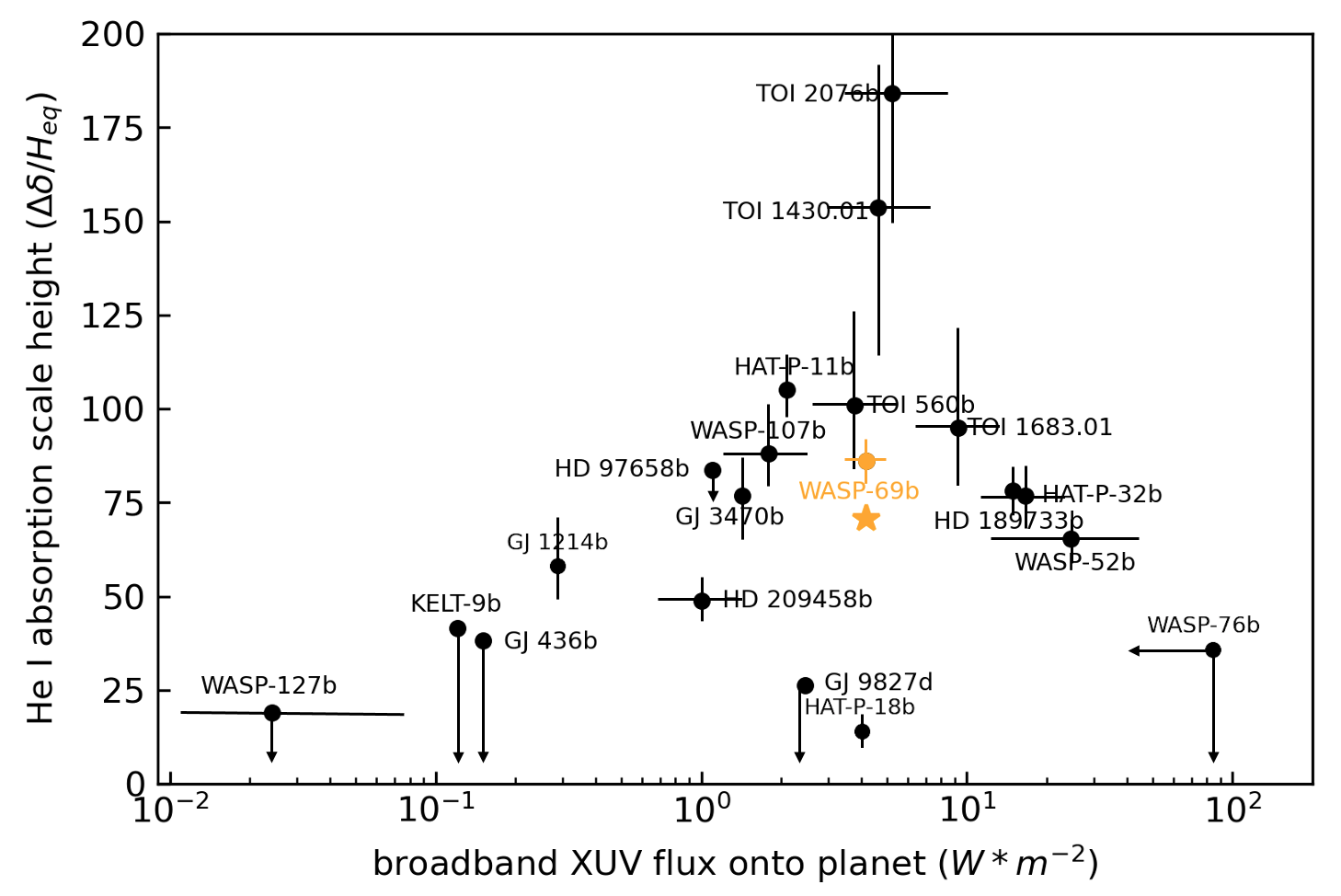}
	\caption{Summary of He~I~10830~\r{A} detections. Equivalent height $\delta_{R_{p}}$ is normalized by the atmospheric scale height $H_{eq}$ and plotted vs. broadband XUV flux the planet receives. The WASP-69b observations are both colored orange, with the results from this work marked with a star. Note that the two WASP-69b data points agree when the transit depth is averaged over the same phase of the transit. Data for other known He~10830~\r{A} detections taken from multiple references\textsuperscript{3}.}
	\label{fig:summary}
\end{figure}

Figure~\ref{fig:summary} shows all of the He~I~10830~\r{A} detections that we are aware of in the literature\footnote{{\cite{2018nova.pres.4320K}; 
 \cite{2018Sci...362.1388N};
 \cite{2018A&A...620A..97S};
 \cite{2019A&A...629A.110A}; 
 \cite{2020MNRAS.495..650G}; 
 \cite{2020ApJ...894...97N};
 \cite{2020AJ....160..258K};
 \cite{2020AJ....159..278V};
 \cite{2020A&A...640A..29D};
 \cite{2021EPSC...15..622C};
 \cite{2021AAS...23754305P};
 \cite{2021AJ....162..284S}; \cite{2021AJ....162..222V};
 \cite{2022AJ....164...24K};
 \cite{2022A&A...659A..55O};
 \cite{2022A&A...657A...6C};
 \cite{2022AJ....163...67Z}; \cite{2023AJ....165...62Z}). }} and shows the two WASP-69b detections plotted in orange (this work indicated with a star). For WASP-69b, using our average absorption over the full transit ($T_{14}$) of 2.7~$\pm$0.4$\%$ we compute $\delta_{R_{p}}$/$H_{eq}$ = 66.1~$\pm$~3.1. We note that our equivalent height to lower atmosphere scale height estimate is about $\sim$ 25$\%$ less than the value 85.5~$\pm$~3.6 reported by \cite{2018Sci...362.1388N} but this is only due to a choice of when to average over the transit chord. \cite{2018Sci...362.1388N} reports the average during $T_{23}$, when the absorption is deeper. Similarly, during $T_{23}$ of our time series, we measure an average absorption depth of 3.6~$\pm$~0.5$\%$ and an equivalent height to lower atmospheric scale height estimate of 83.1~$\pm$~3.9. While we opt to report the standard average absorption over the course of the whole transit ($T_{14}$), we note that these equivalent height values are in agreement.

Here we highlight a practical limit for the equivalent height metric as a mode of comparison between He~I absorption detections. Because of the dependence on transit depth $\Delta d$, instruments with different resolutions may not derive the same equivalent height for the same object. As spectral resolution, $R = \frac{\lambda}{\Delta\lambda}$, increases, spectral features appear narrower and deeper. For lower resolution, the same feature will become broadened and shallow. Thus, the transit depth for the same system can vary across instruments (or with slit width within the same instrument). However, what will remain constant is the total area under the absorption curve. 

We suggest a metric that translates better across observation and instrument parameters such as {\em `equivalent width-time'} $\int\text{EW} dt$ over the full transit/tail duration. In the case of partial transits, an alternative measurement that would make sense is equivalent width per unit orbital phase. This yields a single number that would normalize comparisons across observations/instruments in a convenient way. For our observations, we compute $\int\text{EW} dt = 297 \pm 50$~\r{A}~s. We did not re-analyze other existing datasets to compute this quantity, but such an effort would bring additional clarity into the census of helium outflows.

Although we can confidently attribute a significant portion of the observed variability of this system to differences in SNR between instruments, stellar variability may also contribute. There are at least three known sources of stellar variability that could be causing epoch-dependent variations in the He~I transit depth/morphology/velocity structure. 

The first is that the He~I~10830~\r{A} feature varies in the stellar atmosphere. However, this seems unlikely as we did not measure significant relative He~I absorption variability in any of our out-of-transit observations for WASP-69. 

The second source of stellar variability could come from relative changes in output from different parts of the star's electromagnetic spectrum. Although the 2$^{3}$S metastable state is most efficiently populated by the EUV/FUV flux ratio of K-type stars like WASP-69, the extent of variability of those outputs is not well-constrained. Perhaps the tail is significant but the fluctuating stellar EUV/FUV levels act as a sort of variable light source behind the outflow, adjusting the contrast of what regions of it the observer can detect. As the EUV/FUV flux ratio becomes more favorable for the metastable state, we can see lower regions of column density while higher density regions become optically thick, and vice versa.

A third source of stellar variability could come from variations in the stellar wind strength. Work by \cite{2022ApJ...926..226M} on modeling the extended tail for WASP-107b demonstrates that many of the features we find in our observations: a delayed absorption peak relative to the optical transit mid-point, an asymmetrical He~I light curve, and an accelerating blue shift of gradually decreasing He~I absorption hours after egress. These traits are consistent with interactions of a planetary outflow being suppressed and redirected due to a strong stellar wind. This would indicate variability in the physical length of the tail and could explain why we measured a $\geq$7.5~$R_{p}$ length tail compared to 2.2~$R_{p}$ from \cite{2018Sci...362.1388N} in observations made one year prior. It is possible that all three of the sources of stellar mentioned above are linked to one another; they should all vary with overall variations in stellar activity.

\section{Conclusion}
\label{sec:Conclusion}
In this work, we present high-resolution transmission spectroscopy of the metastable He~I~10830~\r{A} absorption feature for the WASP-69b transit on July 12, 2019 UT. During the transit ($T_{14}$) we detect an average relative helium absorption level of 2.7~$\pm$~0.4\% and a net blue shift of $-5.9$~$\pm$~1.0~km~s$^{-1}$. These values are consistent with the in-transit observations from \cite{2018Sci...362.1388N}. The absorption signal is also consistent with \cite{2020AJ....159..278V}. 

However, we detect continued post-transit absorption of He~I which is not seen in the other observations. This extended absorption lasts for at least 1.28 hours post-transit and never returns to the pre-transit baseline. We set a lower limit for the Helium tail $\geq$7.5~$R_{p}$. 

We attribute most of this variability to the high SNR~per~pixel NIRSPEC achieves, allowing the detection of smaller amounts of He~I and over a longer period of time. While instrumental differences surely play a role in these discrepancies, variations within the star or complicated planetary atmosphere dynamics could also be responsible. 

The asymmetry in the Helium absorption curve is consistent with an outflow being shaped by a strong stellar wind and requires 3D hydrodynamic modeling for the most accurate mass loss estimate. However we estimate a mass loss rate of 1~$M_{\oplus}$~Gyr$^{-1}$ to which we trust within an order of magnitude.

Repeat observations are valuable to probe any variability in the outflow properties. There is likely variability stemming from multiple sources and stellar wind strength and EUV/FUV output variability is not well understood for stars other than our Sun, so planetary outflow observations such as these may a useful method for studying and constraining certain types of stellar variability.

\section*{Acknowledgments}
The helium data presented herein were obtained at the
W. M. Keck Observatory, which is operated as a scientific partnership among the California Institute of Technology, the University of California and the National Aeronautics and Space Administration. The Observatory was made possible by the generous financial support of the W. M. Keck Foundation.

We thank the referee for a thorough review that helped
improve the quality of this work. We thank Lisa Nortmann for providing us with the CARMENES data for WASP-69b which was helpful in comparing results between observations. We also thank Michael Zhang and Fei Dai for useful conversations regarding planetary helium outflows. We'd also like to thank Greg Gilbert for helpful edits of this manuscript. DT is supported in part by the Cota-Robles Fellowship at UCLA, which was instrumental in the advancement of this research.

\bibliography{references}
\end{document}